\documentclass[iop]{emulateapj}

\newcommand{\Mm}{{\mathrm{\, Mm}}}
\newcommand{\kms}{{\mathrm{\, km \; s^{-1}}}}

\newcommand{\secs}{{\mathrm{\, seconds}}}
\newcommand{\mins}{{\mathrm{\, minutes}}}
\newcommand{\G}{{\mathrm{\, G}}}

\shorttitle{Large-Amplitude Longitudinal Oscillations}
\shortauthors{M. Luna et al.}

\begin{document}

\title{On the robustness of the pendulum model for large-amplitude longitudinal oscillations in prominences}

\author{M. Luna\altaffilmark{1,2}, J. Terradas\altaffilmark{3,4}, E. Khomenko\altaffilmark{1,2}, M. Collados\altaffilmark{1,2}, A. de Vicente\altaffilmark{1,2}}

\altaffiltext{1}{Instituto de Astrof{\'{\i}}sica de Canarias, E-38200 La Laguna, Tenerife, Spain}
\altaffiltext{2}{Universidad de La Laguna, Dept. Astrof{\'{\i}}sica, E-38206 La Laguna, Tenerife, Spain}
\altaffiltext{3}{Departament de F{\'i}sica, Universitat de les Illes Balears (UIB), E-07122 Palma de Mallorca, Spain}
\altaffiltext{4}{Institute of Applied Computing \& Community Code (IAC$^3$), UIB, Spain}

\begin{abstract}
Large-amplitude longitudinal oscillations (LALOs) in prominences are spectacular manifestations of the solar activity. In such events
nearby energetic disturbances induce periodic motions on
filaments with displacements comparable to the size of the filaments
themselves and with velocities larger than $20
\kms$. The pendulum model, in which the gravity projected along a rigid magnetic field is the restoring force, was proposed to explain these events. However, it can be objected that in a realistic situation where the magnetic field reacts to the mass motion of the heavy
prominence, the simplified pendulum model could be no
longer valid. We have performed non-linear time-dependent numerical simulations of LALOs considering a dipped magnetic field line structure. In this work we demonstrate that for even relatively weak magnetic fields the pendulum model
works very well.  We therefore validate the pendulum model and show
its robustness, with important implications for prominence seismology
purposes. With this model it is possible to infer the geometry of the
dipped field lines that support the prominence.
\end{abstract}

\section{Introduction}\label{sec:intro}
Large-amplitude oscillations (LAOs) of prominences are recently
attracting more attention in the solar physics community. The number
of reported events is increasing as the current telescope capabilities
allow for continuous full Sun monitoring. LAOs are motions involving
speeds larger than $20 \kms$ according to the classification of
\citet{oliver2002}. These speeds are larger than the local sound speed
in the cool prominence plasma, typically $c_\mathrm{s} \sim 10
\kms$. Thus, LAOs involve supersonic motions and non-linear effects are expected.

LAOs are typically divided in two kinds of motions: longitudinal and
transverse with respect to the spine of the filament \citep[see a
  review by][]{tripathi2009}.  In this work we focus exclusively on
the large-amplitude longitudinal oscillations (LALOs) case. These
motions were first reported by \citet{jing2003} followed by two more
publications by \citet{jing2006} and \citet{vrsnak2007}. In these works the authors reported five LALOs events where the filament plasma move almost parallel to their spines. More recently \citet{zhang2012} reported an oscillation at the limb showing that the motion is along the prominence dipped magnetic field lines. More LALOs observations were reported and characterized in later work
\citep{Li2012,bi2014,shen2014,luna2014}. In all the LALOs reported so far the periods range from 40 to 100 minutes and the amplitudes are between $20 \kms$ to $100 \kms$. In LALOs the prominence
threads move parallel to themselves indicating that the motion is
along the magnetic field. \citet{luna2014} demonstrated that in the
considered event the motion formed $25^\circ$ with respect to the
filament spine. That orientation coincided with the typical observed
orientations of the magnetic field of filaments \citep[see
  e.g.,][]{Leroy1983a,Leroy1984a}. Thus, all above evidences suggest
that LALOs are actually oscillations along the magnetic field.

The nature of LALOs is currently a matter of discussion
since, in principle, several forces could act as the restoring
force. Based on 1D numerical simulations, \citet{luna2012b} proposed
the pendulum model to explain LALOs, where the restoring force is the
gravity projected along the magnetic field lines. These authors
discarded the magnetic origin of the restoring force because the
Lorentz force is always perpendicular to the magnetic field and does
not affect the longitudinal motion. Similarly, the numerical
simulations presented by \citet{luna2012b} show that the gas pressure
gradient force is not important and that the nature of LALOs is not
magnetosonic. These results were confirmed by other 1D simulations
by \citet{zhang2012,Zhang2013a}. \citet{luna2012c} performed an
analytical study of the influence of the magnetic field curvature on
linear longitudinal oscillations. It was found that the longitudinal
oscillations are in general a combination of the pendulum and the slow
modes. However, for typical prominence parameters the contribution of
the slow modes is very small confirming the findings of previous
studies. All these results support the pendulum model for the LALOs.

The main drawback of the 1D modeling is that the magnetic field is
assumed rigid and the motion of the plasma perfectly follows the field
lines. In this regime the magnetic field does not react to the motion
of the plasma. However, the heavy mass of the prominence and the
non-zero value of the plasma-$\beta$ could make the role of the magnetic
field more important and influence the motion of threads
\citep{Li2012}. For example, the mass could change dynamically the
magnetic structure producing non-linear coupling of the longitudinal
and transverse motions and the pendulum model might be no longer
valid. Recently, \citet{terradas2013} studied the different modes of
oscillations of a prominence embedded in a 2D structure in the linear
regime. They used a realistic scenario where the magnetic field reacts
to the plasma motion. The authors found evidences that the
longitudinal motion is affected by the gravity but also by the gas
pressure. However, the authors placed the prominence in a very
high position where the field lines are essentially flat and
convex-downwards. At such position, the longitudinal motion is
strongly influenced by the slow modes as was found in
\citet{luna2012c}.

Due to the potential applicability for prominence seismology, it is necessary to validate or to discard the pendulum
model. In this study we demonstrate that the pendulum model is robust,
with non-linear 2D time-dependent numerical simulations. We prove that
the period of the oscillations depends exclusively on the radius of
curvature of the dipped field lines and that the longitudinal and
transverse motions (relative to the magnetic field) are not strongly coupled. The
back reaction of the magnetic field is small when LALOs are present
and the restoring force is the projection of the gravity along the
magnetic field lines.

\section{The numerical experiment}
The aim of the numerical experiment is to reproduce the LALOs observed
in filaments. In those events nearby energetic disturbances
perturb an existing filament from the side, probably along
the filament-channel magnetic structure. The hot plasma produced at
the energetic event flows along the magnetic field lines pushing an
already formed prominence. In order to have a prominence we
instantaneously load a force free magnetic structure with a prominence
mass. Initially, such a configuration is not in force equilibrium. We
then let the system evolve in order to have a relatively relaxed
configuration. The amplitude of the motions of the system decreases
with time but the velocity field does not disappear completely. Once
we reach a relatively relaxed situation we instantaneously apply a
velocity perturbation along the magnetic field of the prominence in
order to mimic a typical observed excitation event such as a jet or a
small flare. The velocity perturbation is placed over the almost
relaxed prominence with a speed of $20 \kms$, typical for LALOs.

\begin{figure}[!ht]
\begin{center}
\includegraphics[width=0.5\textwidth]{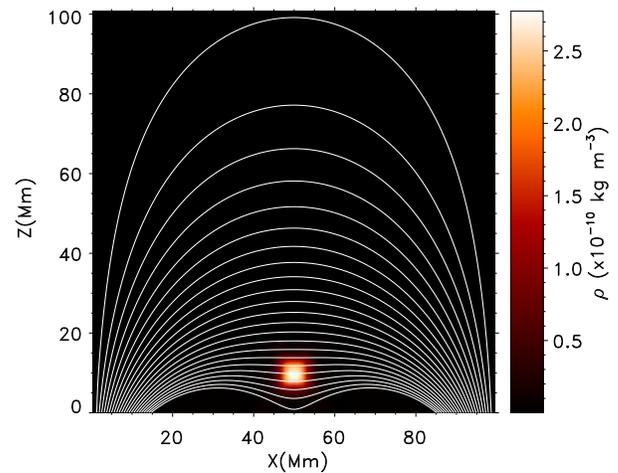}
\end{center}
\caption{Initial configuration consisting of a stratified corona and a force-free magnetic field given by Equations (\ref{eq:fff-conf-x}) and (\ref{eq:fff-conf-z}) \label{fig:initial-conf} and an initial density and temperature prominence perturbation. For the force-free magnetic configuration we have imposed $B_0= 30 \, \mathrm{G}$, $k_1=k_2/3=\pi/100 \Mm^{-1}$, and $z_0=-2 \Mm$. A selected set of field lines have been plotted as white lines. The color field represents the density at the prominence and corona. The density of the internal part of the prominence is $100$ times the density of the corona.} 
\end{figure}

In the relaxation phase, the initial configuration consists of a force-free magnetic field described in \cite{terradas2013}. It represents a symmetric double arcade given by Equations (\ref{eq:fff-conf-x}) and (\ref{eq:fff-conf-z}) and plotted in Figure \ref{fig:initial-conf}. 
\begin{eqnarray}\label{eq:fff-conf-x}
\frac{B_x(x,z)}{B_0}&=&\cos k_1 x \, e^{-k_1 (z-z_0)}-\cos k_2 x \, e^{-k_2 (z-z_0)} ~,\\ \label{eq:fff-conf-z}
\frac{B_z(x,z)}{B_0}&=&- \sin k_1 x \, e^{-k_1 (z-z_0)}+ \sin k_2 x \, e^{-k_2 (z-z_0)}~.
\end{eqnarray}

This structure has dips closer to the surface, where we initially
place the prominence mass. The depth of the dips decreases
monotonically from bottom to top positions and the curvature at the
central position of the field lines changes from concave upwards to
convex downwards forming a overlying arcade of loops.

The magnetic field has a null point at $(0,z_0)$ that we have placed outside the numerical domain by taking $z_0=-2 \Mm$. In addition, placing the null point bellow the numerical domain we avoid partially the non realistic region of very high plasma-$\beta$ (ratio of gas pressure to magnetic pressure) surrounding the null. Both pressures change dynamically with time, but the plasma-$\beta$ is always constrained to $\beta < 0.4$ in all the numerical domain. We have considered the parameters $B_0 = 30 \G$ and $k_1=k_2/3=\pi/100 \Mm^{-1}$. At the central position of the structure the magnetic field strength is $3.3 \G$ at $z=0$ and reaches a maximum of $11.5 \G$ at $z= 15.5 \Mm$. Beyond this point the field strength decreases with height. This point coincides with the change of curvature of the field lines. Thus, below $z=15.5 \Mm$ the magnetic structure has dips whereas above this point the structure is an arcade of loops. The numerical domain consists of a box of $800\times800$ points and $100\times100 \Mm$, with a spatial resolution of 125 km. We assume open boundary conditions for the top and the two side boundaries. At the bottom boundary, line-tying is considered by imposing current free conditions and zero velocities as in \citet{terradas2013}.

The coronal plasma is assumed initially stratified with a uniform million degree temperature, $T_c = 10^6 \, \mathrm{K}$. The prominence is placed on the structure by perturbing the coronal atmosphere with a temperature of the form
\begin{equation}\label{eq:temp_pert}
T(x,z)=T_c + \delta T_0 \, e^{ -\frac{(x-x_T)^4}{w_x^4} -\frac{(z-z_T)^4}{w_z^4} }
\end{equation}
where $(x_T,z_T) = (50,10) \Mm$ is the position of the center of the pulse and $w_x=w_z=10 \Mm$ are the widths of the perturbation in $x$- and $z$-directions respectively. The 
temperature at the central position of the prominence is $T = 10.000 \, \mathrm{K}$ and $\delta T_0$ is selected accordingly to $-0.99 \times 10^6 \, \mathrm{K}$. The initial temperature perturbation has a quartic dependence in the exponent in order to have a large region of cool and dense plasma. The pressure is unperturbed and equal to that of the ambient corona. The density of the perturbation is computed accordingly to the ideal gas law (see Fig. \ref{fig:initial-conf}). In this situation, the gas pressure along the horizontal direction is uniform, $\rho_\mathrm{c}(z) \, T_\mathrm{c} (z) = \rho_\mathrm{p}(x,z) \, T_\mathrm{p}(x,z)$.

We numerically solve the ideal MHD equations in the above scenario by
using our recently developed MANCHA code
\citep{khomenko2008,felipe2010,khomenko2012,khomenko2014}. MANCHA
solves the time dependent nonlinear problem given initial
conditions. The relaxation phase is not relevant for this study, but
it is sufficiently interesting to be briefly described. The initial
configuration is not in equilibrium and the weight of the prominence
mass is not compensated by the magnetic structure or gas pressure
gradients. The prominence plasma rapidly starts to fall and collapses, increasing its internal pressure. After some time the collapse ceases producing a
shock wave that propagates along the magnetic field lines at both sides of
the prominence. These shocks bounce at the field footpoints moving
back again to the prominence. The shocks travel between the two
footpoints, and the prominence oscillates vertically producing again
more shocks by the same process. After $7000 \secs$ the prominence is
more or less relaxed into a dynamical equilibrium. At this point vertical
oscillations and waves along the magnetic field remain but the speeds involved are
below $5 \kms$. It is important to note that we include the remnant
velocities in the next phase where we add the longitudinal
perturbation to the already existing velocity field. The resulting
dynamics will not be strongly influenced by these remnant motions.

\section{Large-Amplitude Longitudinal Perturbation and Oscillations}\label{sec:lalos}
Once the prominence is sufficiently relaxed we perturb the resulting configuration with a velocity field along the magnetic structure, namely
\begin{equation}\label{eq:initial_vel}
\vec{v}_0= v_0 \frac{\vec{B}_0}{B_0}  e^{-\frac{(x-x_T)^4}{w_{v x}^4} -\frac{(z-z_T)^4}{w_{v z}^4}} \,, 
\end{equation}
in order to excite longitudinal motions. We have considered this time as $t=0$. The relaxation phase has $t<0$, while $t \ge 0$ is taken after the longitudinal velocity perturbation. $\vec{B}_0$ is the magnetic field at the end of the relation phase or at $t=0$. The perturbation has $w_{v x} = 3 \, w_{x}$ and $w_{v z} = 3 \, w_{z}$, much larger than the initial temperature perturbation in order to excite oscillations of the cool plasma and in their surrounding. The perturbed initial velocity is set to $v_0= 20 \kms$ which is in the range of large-amplitude oscillations. Due to the dimensions of the considered structure, larger initial velocities produce drainage of cool plasma on the sides of the structure.

\begin{figure*}[!ht]
\includegraphics[width=0.99\textwidth]{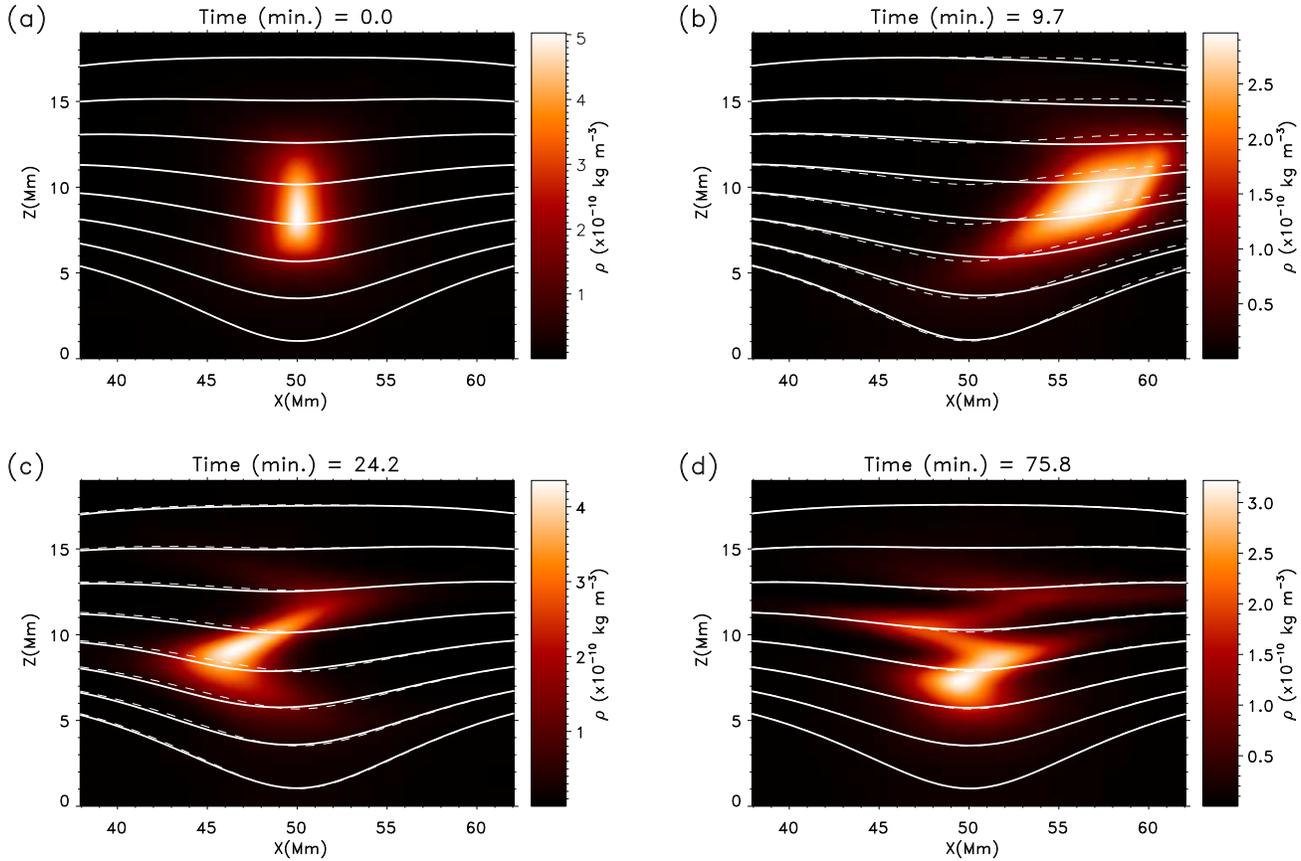}
\caption{Time sequence evolution of the plasma (colors) and the magnetic structure (white lines) at four times. Panel (a) shows the initial configuration at $t=0$ when the longitudinal perturbation is applied. In (b) we see the prominence almost reaching the maximum elongation of its displacement. In (c) the phase differences between different parts of the structure is clear. In (d) a clear zig-zag motion of the prominence is shown. The color represents the density of the prominence.}\label{fig:time-sequence-lalo}
\end{figure*}

In Figure \ref{fig:time-sequence-lalo} we have plotted a four panel time sequence of the density and the magnetic field. 
In Figure \ref{fig:time-sequence-lalo}a the almost relaxed prominence
is plotted. In contrast with the non-equilibrium initial situation
(Fig. \ref{fig:initial-conf}) the density is now more vertically
elongated because the cool plasma has deformed the initial magnetic field. We perturb the system with the velocity of Equation
(\ref{eq:initial_vel}) and the whole prominence material starts to move towards
the increasing values of $x$-coordinate following clearly the magnetic
field lines. The velocity also decreases with time and after
approximately $10 \mins$ most of the prominence reaches the maximum
displacement (Fig. \ref{fig:time-sequence-lalo}b) and the motion
reverses. However, the bottom parts of the prominence reverse the
direction of motion slightly before the rest of the
structure. The motion along the vertical structure is different and
the phase difference becomes important, as it is apparent in Figure
\ref{fig:time-sequence-lalo}c. At this time the bottom part moves
rightward and the central part leftwards with the top part
considerably delayed. The increase in the delay between the different parts of the structure seems gradual. The last panel
(Fig. \ref{fig:time-sequence-lalo}d) shows very well that the motion
between different parts of the structure becomes out of phase with
opposite directions of motion. This indicates that the plasma moves
along the magnetic field with different periods of oscillation. The
zig-zag structure resembles the motion of the apparent tornado structure observed with the SDO/AIA, as reported by
\citet{su2012}. From the figure it is clear that the density changes
with time in reaction to both longitudinal and transverse movements.

The magnetic field structure also changes during the temporal evolution. In the figure, the white solid lines are a set of selected field lines at different times and the white dashed lines represent the magnetic field at $t=0$ (Fig. \ref{fig:time-sequence-lalo}a). In the first period the magnetic field changes slightly (Figs. \ref{fig:time-sequence-lalo}b and \ref{fig:time-sequence-lalo}c) but rapidly recovers the initial configuration and remains more or less unchanged for the rest of the temporal evolution (see Fig. \ref{fig:time-sequence-lalo}d). In the time-dependent evolution we do not find a swaying motion of the magnetic structure. In contrast, the magnetic structure remains more or less unchanged except for some small-amplitude vertical oscillations. We conclude that the motion of the plasma mainly follows the magnetic field lines.

We see that different parts of the structure oscillate with
different frequencies forming a continuous spectrum
\citep{goossens1985,terradas2013}. This behavior makes the spectral
study of the motion complicated. In addition, these oscillations are
essentially non-linear and the plasma is subject to important
advection. In our fluid, the frozen-in condition applies because we
consider it to be perfectly conducting. Thus, it is necessary to
advect every field line considered in order to catch the
longitudinal and transverse motions of the plasma. The bottom
line-tying boundary conditions imply that the magnetic field is
unperturbed during the temporal evolution. Thus, any field line
starting at the bottom boundary follows the plasma motion simplifying
the study of the oscillations. We select a large set of equally spaced field lines
starting at the bottom. These field lines
cross the central axis ($x=0$) with approximate separations of $0.25
\Mm$.  The lines plotted in Figure \ref{fig:time-sequence-lalo} are only a small subset of the field lines considered in the following analysis.

\begin{figure*}[!ht]
\begin{center}
\includegraphics[width=0.9\textwidth]{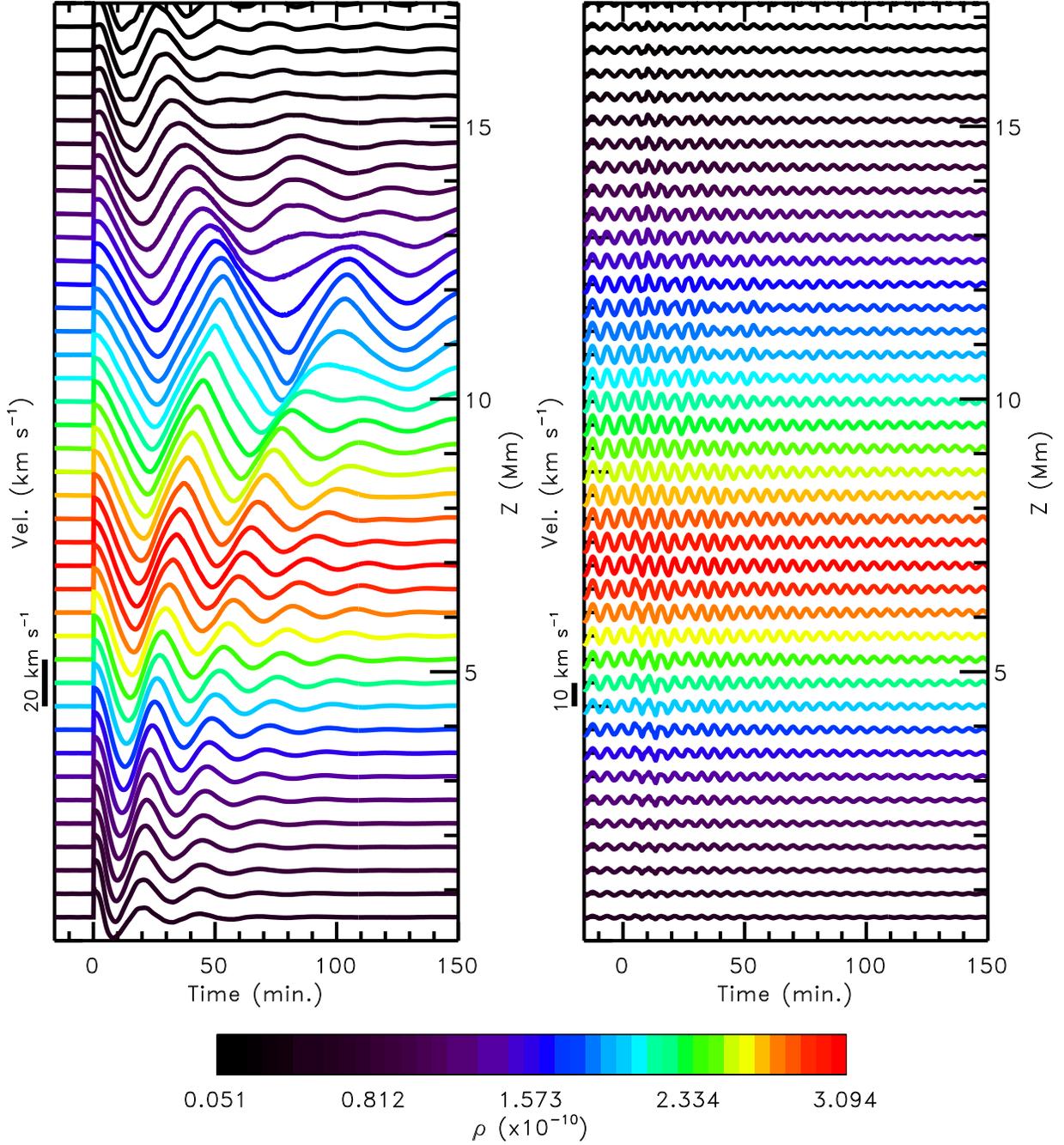}
\caption{Plot of the longitudinal (left panel) and transverse (right panel) motions of the center of mass for every field line considered. In this plot the vertical axis represents an arbitrary scale. On each panel a velocity scale is plotted. This plot is  convenient to show the relative oscillation phase at the different heights of the structure.}\label{fig:lalo-velocities}
\end{center}
\end{figure*}

We have computed the longitudinal and transverse motions of the center of mass as 
\begin{eqnarray}
v_{\parallel i}(t)= \int v_\parallel(s_i,t)  \rho(s_i,t) ds_i / \int \rho(s_i,t) ds_i \,, \\
v_{\perp i} (t) = \int v_\perp(s_i,t)  \rho(s_i,t) ds_i / \int \rho(s_i,t) ds_i \,. 
\end{eqnarray}
The $i$-index corresponds to the field line considered and $s_i$ is the coordinate along the field line $i$ that depends implicitly on the time, $t$. The integrands $v_\parallel(s_i,t)$, $v_\perp(s_i,t)$ and $\rho(s_i,t)$ are the longitudinal velocity, transverse velocity and density along the field line $i$ respectively. These velocities are plotted in Figure \ref{fig:lalo-velocities} for the different field lines.
The left vertical axis shows the velocity scale. For comparison, the right vertical axis shows the $z_i$-coordinate at the central position $x=0$ of the field lines at $t=0$. In a linear situation this is equivalent to studying the velocities along the vertical axis at $x=0$. We have plotted the temporal evolution of the velocities from $t=-15 \mins$ to $t = 150 \min$. Before $t=0$ we clearly see the velocities of the relaxation phase at the right panel. At the relaxation phase the central positions of the prominence oscillate vertically because we place the mass centered at the structure. We see that the longitudinal velocities are zero (left panel) but there are clear oscillations along the vertical direction (right panel). At $t=0$ we impose the instantaneous velocity perturbation that produces a sudden increase of the longitudinal velocity to $20 \kms$, but no significant changes to the transverse motion. The different nature of the longitudinal and transverse oscillations is also clear in their very different periods of oscillation. We define the density at the center of mass of each field line as ${\rho_{CM}}_i (t) = \rho({s_{CM}}_i,t)$ being ${s_{CM}}_i$ the position of the center of mass along the field line $i$. The color coding is computed as the temporal average of ${\rho_{CM}}_i (t)$. With this color coding we can identify the oscillation corresponding to coronal or prominence plasma. The longitudinal period increases monotonically from the bottom positions up to $z \sim 11 \Mm$. This range of heights includes almost all the prominence. Above $z \sim 11 \Mm$ the period of the oscillations decreases. From the figure we see that the longitudinal velocity at denser plasma shows a triangular shape with peaked maximums and minimums instead of a sinusoidal shape. This could be related to the non-linearity of the oscillations. 

The longitudinal oscillations show a significant damping (Fig. \ref{fig:lalo-velocities}a) in contrast to the weak damping of the transverse motions (Fig. \ref{fig:lalo-velocities}b). LALOs show different damping times for different parts of the structure. The plasma oscillations at the bottom parts of the structure are damped quicker than at the top parts. In order to study the convergence of the numerical solutions, we have performed a check repeating our simulations with a decreased resolution of 500 and 250 km grid sizes. We found that at all resolutions the simulations show a similar damping scale, with slightly larger damping for larger grid sizes and identical periods. Therefore, these tests demonstrate that our simulations have converged. We may speculate that a kind of numerical phase-mixing is probably occurring for LAL oscillations, mimicking a real process that may occur on the Sun. LALOs at different field lines have different frequencies because they have different field line curvature. Then two adjacent field lines at the numerical grid lose their relative coherence and due to the numerical viscosity the motion tends to be canceled by stress. The numerical viscosity depends on the grid size. However, the convergence test has shown that we are very far from reaching realistic values of dissipation because the models with different resolution exhibit similar damping. Thus, significantly better grids are required to resolve the transition region surrounding the oscillating thread and to avoid the enhanced numerical viscosity phase-mixing. This will be addressed in a future work using a new module of MANCHA with Adaptive Mesh Refinement. The study of the damping mechanism is out of the scope of this work. However, understanding the nature of this damping is also important for LALOs and it will be considered in a future publication. 

The transverse velocities in Figure \ref{fig:lalo-velocities} (right panel) show an amazingly regular and coherent motion. After the perturbation ($t > 0$), there is an alteration of the regular pattern, but the system quickly recovers its very regular motion. There are no important phase differences, and the oscillation periods are very similar ($5.4 \mins$ in all field lines analyzed). This behavior suggests that the transverse motions is a global normal mode of the structure. However, one of the most important results is that there are no transverse motions associated to the longitudinal motion.The back reaction of the magnetic field should have produced displacements perpendicular to the magnetic field with the same or half period than the longitudinal motion. However, this is not observed in Figure \ref{fig:lalo-velocities} or in the spectral analysis of the curves.

\begin{figure}[!ht]
\begin{center}
\includegraphics[width=0.5\textwidth]{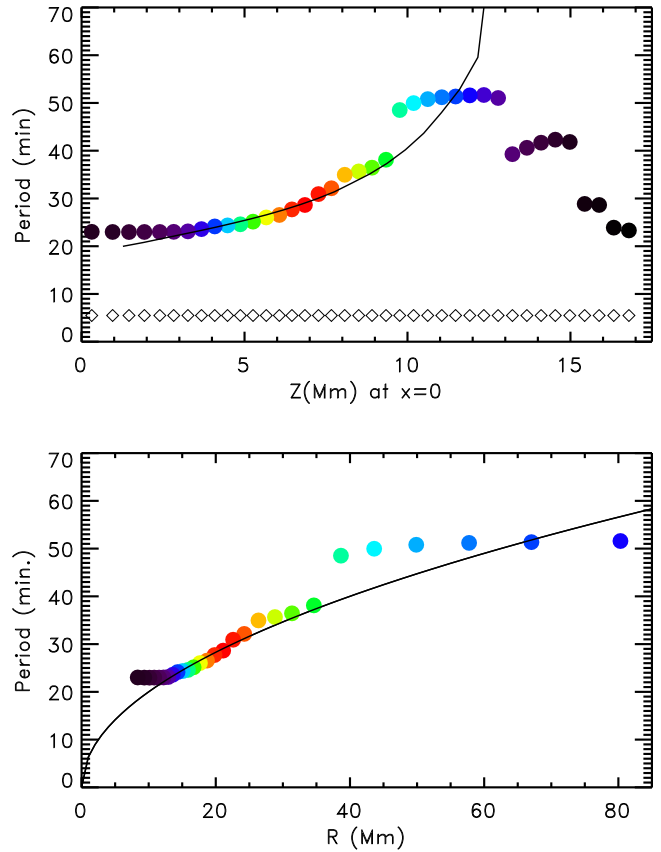}
\caption{Scatter plot of the periods of oscillation as function of $z$-position of the field lines dips, as in Figure \ref{fig:lalo-velocities} (top panel), and radius of curvature (bottom panel) of the field lines. The radius of curvature is computed numerically along the field lines and averaged in space and time. The radius of curvature for every field line is spatially averaged in $3 \Mm$ along the line centered at the dip for every snapshot. This spatially averaged radii are also temporally averaged in all the simulation time $t \ge 0$. The color code is identical to Fig. \ref{fig:lalo-velocities}.}\label{fig:lalo-periods}
\end{center}
\end{figure}

We have analyzed the periods of oscillations at every field line considered in the two polarization directions. The results are plotted in Figure \ref{fig:lalo-periods}.
The longitudinal period increases with height (top panel) as we have discussed previously. The increase is monotonic but above $z \sim 9.5 \Mm$ the behavior changes, in agreement with the visual impression from Figure \ref{fig:lalo-velocities}. In the figure we have plotted the theoretical period, $P$, of the mass loaded into the deformed structure (solid line) using the pendulum model expression from \citet{luna2012b},
\begin{equation}
P=2 \pi \sqrt{\frac{R}{g}} \, ,
\end{equation}
where $R$ is the radius of curvature of the dipped portion of the field lines and $g=274\, \mathrm{m \, s^{-2}}$ is the solar gravity acceleration. We have computed the radius of the curvature at the center of the dipped portion of the field lines. These radii slightly change in reaction to the mass motion. We therefore have considered the temporal average of the radii. We can see an excellent agreement between the time-dependent simulations and the pendulum model in the range $z = 3-9.5 \Mm$ that includes most of the prominence mass. Above $z \sim 9.5 \Mm$ there is no agreement with the model. Above this point the gas pressure gradient contributes largely to the restoring force. The reasons are that the field lines supporting the cool plasma are flatter with small projection of the gravity along the field and the density contrast is relatively small. In this situation the gas pressure gradient becomes important \cite[see][]{luna2012c}. For this reason, above $z=9.5 \Mm$ the frequencies do not fit the pendulum curve. In this work we are considering field lines of the order of $100 \Mm$. This length is probably in the range of short prominence field lines. In typical prominences the field lines supporting the cool plasma are larger. In \cite{luna2012c} we found that for longer magnetic field lines the pendulum works even better. Similarly, below $z = 3 \Mm$ the medium is essentially coronal plasma and the oscillation modes are slow modes. In the same figure we have plotted the period of the transverse oscillation as diamonds (top panel). The periods are uniform with a common value of $5.4$ minutes. The uniformity of the period suggests that the transverse motion is related to the fast wave global normal mode. 
Figure \ref{fig:lalo-periods} (bottom) shows a very important result for prominence seismology with profound observational implications. In this figure we have plotted the period as a function of the radius of curvature of the field lines. This clearly demonstrates that the pendulum model of \citet{luna2012b} works exceedingly well for LALOs in prominences. Thus, the period of oscillation of the LALOs is only dependent on the radius of curvature of the dipped field lines that support the plasma. This very good match between the numerical simulations and analytic expressions shows that the plasma contained at different heights is oscillating along the magnetic field and gravity is the restoring force. The zig-zag shape of the oscillating prominence is entirely associated to the different periods of oscillation along the vertical structure that produces phase differences between the motion of the different flux-tubes.

\section{conclusions}
In this work we have performed non-linear time-dependent numerical simulations of LALOs considering a dipped magnetic field line structure. The motion is very complex but we have not found evidences of strong coupling between longitudinal and transverse motions of the plasma. This indicates that the longitudinal motion of the plasma along the magnetic structure does not produce significant vertical motions of the field due to back reaction.

We have also demonstrated that LALOs are very well described by the pendulum model. The periods of the numerical simulations agree very well with the results of the analytical model. This demonstrates that there is one to one relation between the period of the oscillations and the radius of curvature of the field lines. This result has important observational implications because it is possible to easily infer the geometry of the filaments when LALOs are present in such structures.

\acknowledgments
M. Luna, E. Khomenko, M. Collados and A. de Vicente acknowledge the support by the Spanish Ministry of Economy and Competitiveness through projects AYA2011-24808, AYA2010-18029 and AYA2014-55078-P. This work contributes to the deliverables identified in FP7 European Research Council grant agreement 277829, ``Magnetic Connectivity through the Solar Partially Ionized Atmosphere'' (PI: E. Khomenko). J. T. acknowledges support from the Spanish ``Ministerio de Educaci{\'o}n y Ciencia'' through a Ram{\'o}n y Cajal grant and support from MINECO and FEDER funds through project AYA2014-54485-P. M. L. and J. T. acknowledge support from the International Space Science Institute (ISSI) to the Team 314 on ``Large-Amplitude Oscillation in prominences'' led by M. Luna. The authors acknowledge the partial contribution of Teide High-Performance Computing facilities to the results of this research. TeideHPC facilities are provided by the ``Instituto Tecnol{\'o}gico y de Energ\'
{\i}as Renovables'' (ITER, SA). URL: \url{http://teidehpc.iter.es}. Resources partially supporting this work were provided by the NASA High-End Computing (HEC) Program through the NASA Center for Climate Simulation (NCCS) at Goddard Space Flight Center.


\end{document}